\documentclass{PoS}
\usepackage{amsmath}                                                                                 

\title{The $\mathbb{C}$P(2) Model at Non-Zero Chemical Potential}

\ShortTitle{The $\mathbb{C}$P(2) Model at Non-Zero Chemical Potential}

\author{\speaker{Wynne Evans$^\dagger$}, Urs Gerber$^{\ a,b}$, Uwe-Jens Wiese$^\dagger$\\ 

  $^\dagger$Albert Einstein Center for Fundamental Physics, Institute for Theoretical Physics,\\
  \ University of Bern, Sidlerstrasse 5, CH-3012 Bern, Switzerland.\\

  $^a$Instituto de Ciencias Nucleares, Universidad Nacional Aut\'onoma de M\'exico,\\ 
  \ A.P. 70-543, C.P. 04510 Ciudad de M\'exico, Mexico.\\

  $^b$Instituto de F\'isica y Matem\'aticas, Universidad Michoacana de San Nicol\'as de Hidalgo,\\ 
  \ Edificio C-3, Apdo. Postal 2-82, C.P. 58040, Morelia, Michoac\'an, Mexico.\\

  E-mail: \email{evans@itp.unibe.ch}}

\abstract{Recently the simulation of quantum field theories using man-made physical systems has become realistic. In this publication we present numerical results which support the use of quantum simulation experiments to study quantum field theories at non-zero chemical potential. We have numerically simulated the (1+1)-d $\mathbb{C}$P(2) model, which shares several interesting features with QCD, namely asymptotic freedom, a dynamically generated mass gap and topological sectors, via dimensional reduction of a (2+1)-d microscopic theory of SU(3) quantum spins. Numerical results for the particle number density as a function of chemical potential are presented.}

\FullConference{34th annual International Symposium on Lattice Field Theory\\
		24-30 July 2016\\
		University of Southampton, UK}

\begin{document}

\section{Introduction}
\vspace{-8pt}
A major contemporay goal of theoretical physicists is to study QCD at non-zero chemical potential. Recently the simulation of quantum field theories using man-made physical systems has become a real possibility \cite{Nascimbene12}. In \cite{Laflamme:2015wma,Laflamme:2015pos} an experimental set-up to quantum simulate the $\mathbf{C}$P(2) model using alkaline-earth atoms trapped in an optical lattice was proposed. The theoretical model for the proposed experiment is an SU(3) quantum spin ladder. Using Monte Carlo techniques, the proposal was supported by numerically simulating this quantum spin system. In this publication we present numerical results which support the use of quantum simulation experiments to study quantum field theories at non-zero chemical potential. 

We simulate the (1+1)-d $\mathbb{C}$P(2) model, which shares several interesting features with QCD, namely asymptotic freedom, a dynamically generated mass gap and topological sectors, via dimensional reduction of a (2+1)-d microscopic theory -- an SU(3) quantum spin ladder. We present numerical results for the particle number density as a function of chemical potential, which confirm that our theoretical framework is robust.

\vspace{-8pt}
\section{Theoretical Framework}
\vspace{-8pt}

The theoretical framework is that of D-theory \cite{Brower:1999kq,Brower2004149,Beard:2004jr}, in which the target field theory emerges from the {\it dimensional} reduction of a system of {\it discrete} variables. The microscopic theory of discrete variables is chosen such that at low energies its global symmetry spontaneously breaks to give a certain number of Nambu-Goldstone bosons, which emerge as degrees of freedom in the coset space of the global broken and unbroken symmetry groups.

\begin{figure}
\centering
\includegraphics[scale=0.37, trim = 0 60 0 630, clip=true]{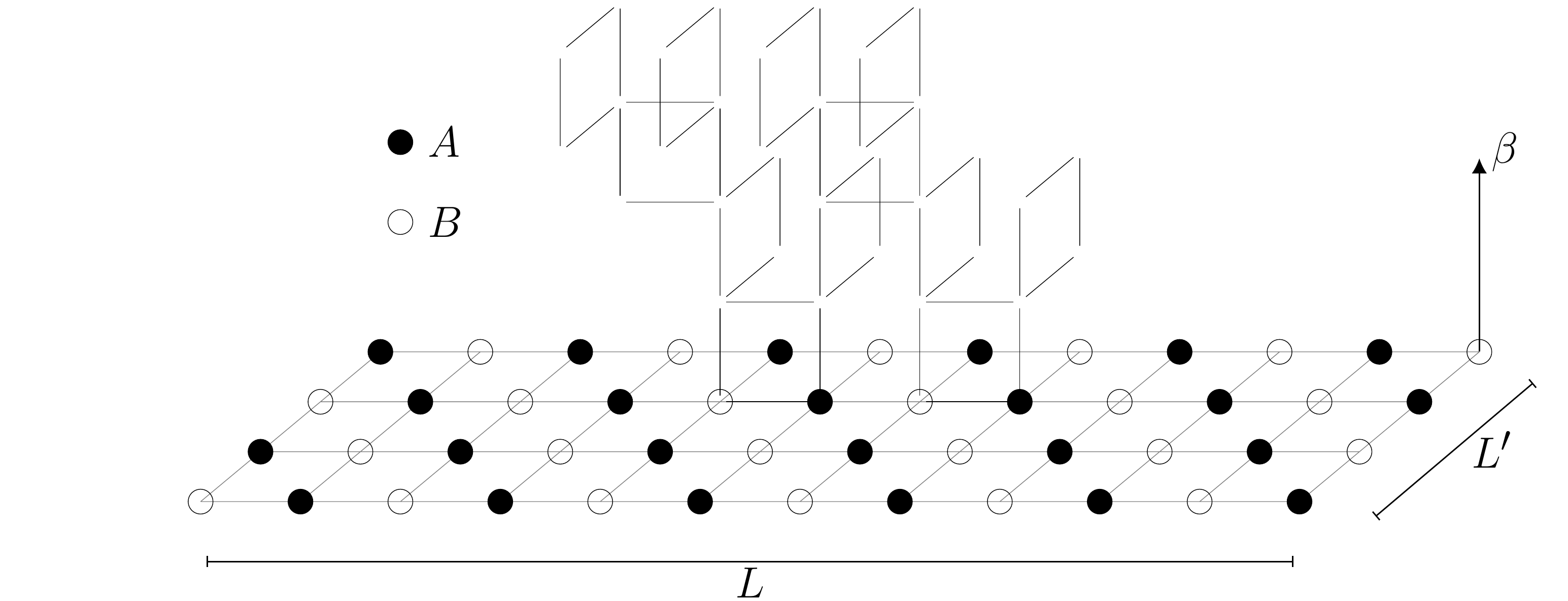}
\caption{The quantum spin variables are positioned in a periodic, square, bipartite lattice; spins living on sublattice $A$ and $B$ transform under the fundamental and antifundamental representations of SU(3), respectively. The transparent plaquettes extending into the $\beta$ direction illustrate the Trotter decomposition of a single Euclidean time step $\epsilon$.}
\label{fig:lattice}
\end{figure}

Since the (1+1)-d $\mathbb{C}$P(2) model has a global SU(3) symmetry, we investigated a microscopic theory with a nearest-neighbour interaction between SU(3) quantum spins located at a site $x$ on an $L\times L'$ periodic, square, and bipartite lattice with spacing $a$, as shown in Figure \ref{fig:lattice}. The Hamiltonian is,
\begin{equation}
H=-J\sum_{\substack{\left<xy\right>\\x\in A}}T_x^aT_y^{a*} - \mu^aT^a, \ \ \  T^a =\sum_{x\in A}T^a_x - \sum_{y\in B}T^{a*}_y,
\label{eq:ham}
\end{equation}
where the SU(3) quantum spins $T^a_x=\frac{1}{2}\lambda^a_x$, obey the usual commutation relations, $[T^a_x,T^b_y]=i\delta_{xy}f_{abc}T^c_x$, and $\lambda^a$ and $f_{abc}$ are the generators and structure constants of SU(3), respectively. $\mu^a$ is a vector of the chemical potentials in the directions $\lambda^a$, but we need only include $\mu_3$ and $\mu_8$ since for SU(3) two independent Casimir operators can be constructed. Also, we have set $\hbar=1$. 

Spins living on sublattice A transform in the fundamental representation \{3\}, while those living on sublattice B transform in the antifundamental representation $\{\bar{3}\}$, which is written explicitly as $-T_x^{a*}$ rather than $\bar{T}_x^a$. For $\mu^a=0$, the system has a global SU(3) symmetry and hence a total spin conservation, $[H,T^a]=0$. An antiferromagnetic coupling, $J>0$, and the fundamental-antifundamental nature of the nearest neighbour interaction are chosen to obtain the relativistic spectrum required for $\mathbb{C}$P(2) model physics. 

This choice also means that in the thermodynamic limit, $\beta,L,L'\rightarrow\infty$, the global SU(3) symmetry breaks to U(2) \cite{Sachdev89, Harada:2003}, giving rise to four\footnote{SU(3) has eight generators, U(2) has four, hence this symmetry breaking results in $8-4=4$ Goldstone bosons.} massless Nambu-Goldstone bosons that live in the coset space SU(3)$/$U(2)$ \ =\mathbb{C}$P(2). The emergent fields are conveniently described by a $3\times 3$ Hermitean projection matrix field $P$, with $P=P^2$, Tr\ \!$P=1$, $P=P^{\dagger}$, and by the (2+1)-d effective action,
\begin{equation}
S[P]= \int_0^\beta \! dt \int_0^L \! dx \int_0^{L'} \! dy \,  {\rm Tr}\big[\rho_s \partial_i P\partial_i P +\frac{\rho_s }{c^2}
D_tPD_t P \big],
\end{equation}
where $\rho_s$ and $c$ are the spin stiffness and spinwave velocity, respectively, $D_tP=\partial_tP-[\mu^aT^a,P]$ rather than $\partial_tP$ includes the chemical potential contribution, and the index $i$ is for an implicit summation over the two spatial dimensions.

Now, if $L'$ is made finite the excitations of the emergent fields only {\it feel} two infinite dimensions. The Mermin-Wagner theorem then insists that the bosons pick up a nonperturbatively generated mass gap, $m=1/\xi$ where $\xi$ is the temporal correlation length. Since the (1+1)-d $\mathbb{C}$P(2) model is asymptotically free and $\xi\propto{\rm exp}[4\pi L'\rho_s/{3c}]$,  then if $L'$ is chosen such that $\xi \gg L'$ the (2+1)-d system will undergo dimensional reduction and moreover the continuum limit will be approached. The dynamics of the resulting system are described by the (1+1)-d effective action,
\begin{equation}
S[P]=\frac{c}{g^2}\int_0^\beta\! dt\!\int_0^L\!\!\!dx\ {\rm Tr}\left[\partial_xP\partial_xP+\frac{1}{c^2}D_tPD_tP\right]
\end{equation}
where $g^2=c/\rho_sL'$ is the dimensionless coupling constant of the dimensionally reduced theory.

\begin{figure}
\centering
\includegraphics[scale=0.4, trim = 0 60 0 60, clip=true]{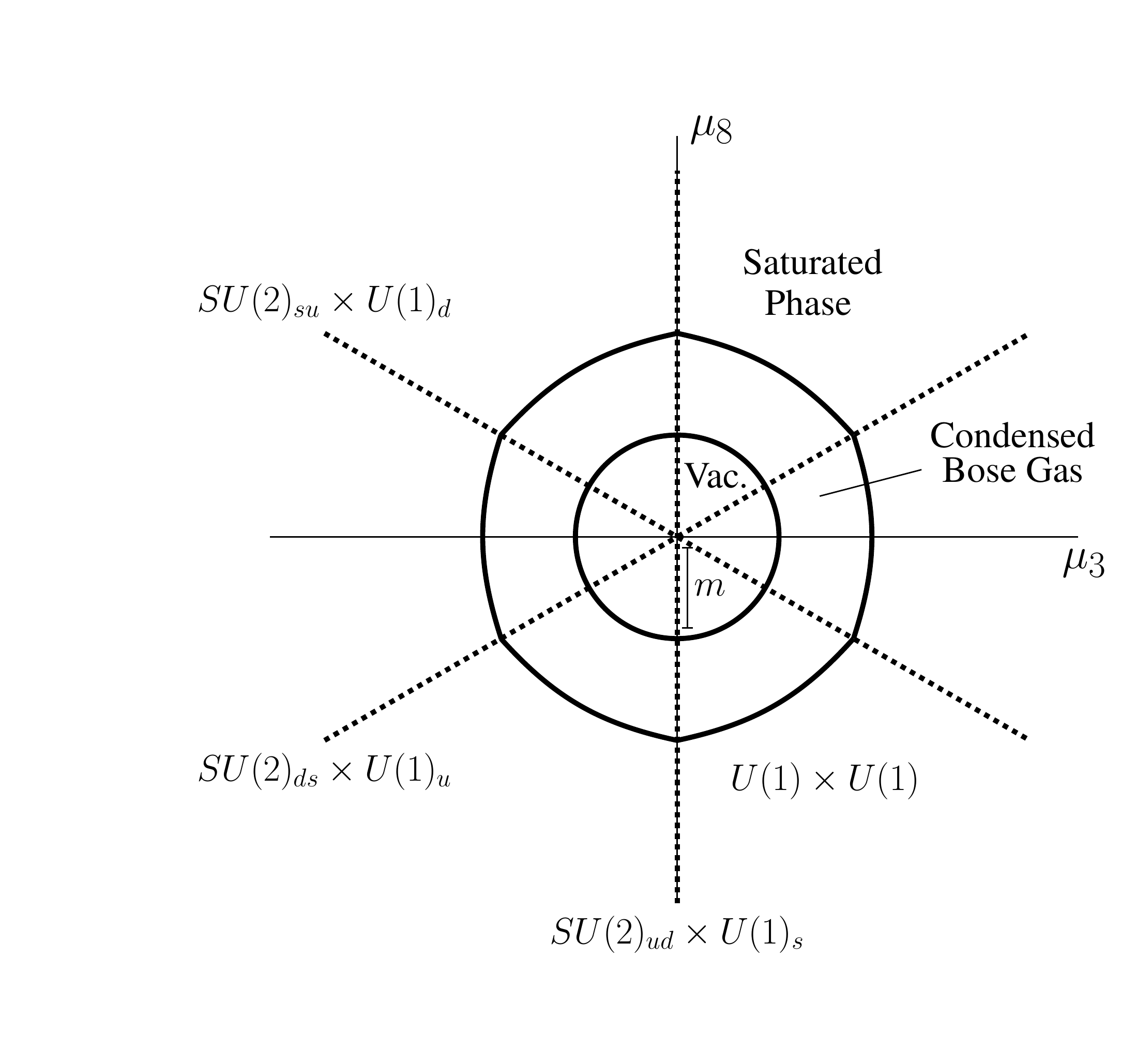}
\caption{A schematic picture of the conjectured $\mu_3$-$\mu_8$ phase diagram for the lattice  $\mathbb{C}$P(2) model. At $\mu_3\!=\!\mu_8\!=\!0$ the $\mathbb{C}$P(2) model enjoys a global SU(3) symmetry, away from this point the global symmetry reduces to U(1)$\times$U(1) except along the dashed lines, where $\lambda_3=\pm\sqrt{3}\lambda_8$ and the global symmetry is SU(2)$\times$U(1). For $\mu<mc^2$, where $m$ is the mass gap and $\mu=\sqrt{ \mu_3^2 + \mu_8^2 }$, we expect a region of vacuum, while for large values of $\mu$ we expect to reach a state of saturation -- all spins aligned . In the intermediate region, $\mu>mc^2$ but not large, the phase of the system is a matter of investigation, but we expect the produced bosons to undergo some form of condensation.}
\label{fig:cp2pd}
\end{figure}
\vspace{-8pt}
\section{Numerical Results}
\vspace{-8pt}
Using Monte Carlo techniques, a numerical simulation of the SU(3) quantum spin system described by \eqref{eq:ham} has been performed. A Trotter decomposition was employed to discretize the time direction such that $\beta=N_t\epsilon/4$, where $N_t$ is the number of lattice points in the temporal direction and $\epsilon$ is a Euclidean time step.  Figure \ref{fig:lattice} illustrates the plaquette break-up resulting from the Trotter decomposition procedure for a single Euclidean time step. 

In \cite{Beard:2004jr} and \cite{Beard:2006mq} a meron-cluster algorithm was used to solve the sign problem at non-zero chemical potential in the $\mathbb{C}$P(1) model, here the $\mathbb{C}$P(2) model is updated using a worm algorithm, which is capable of updating the system at non-zero chemical potential without encountering a sign problem. The {\it worm rules} were derived using information provided by \cite{Evertz2003,Sandvik2002}. In this proceeding we only show results for lattices with periodic boundary conditions and we always keep $\beta c \approx L$, where a value of $c=1.7763(2)Ja$ was determined using the procedure exhibited in \cite{Jiang:2010cd}.

\begin{figure}
\centering
\includegraphics[scale=0.8, trim = 0 0 0 0, clip=true]{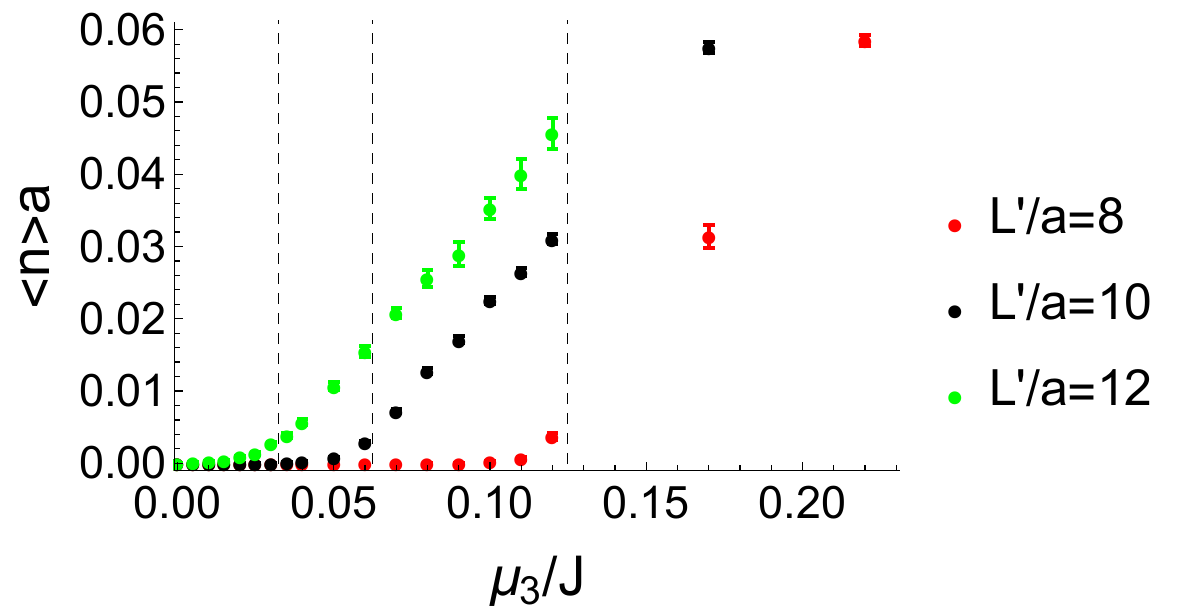}
\caption{Particle number density plotted against chemical potential for $L/a=250$, $\beta J=140.75$ and $\epsilon J=0.05$ for $L'/a=8,10,12$. The dashed vertical lines mark the rest energy gaps calculated for each case: 0.1248(10)$J$, 0.0623(2)$J$, and 0.0322(3)$J$, respectively. }
\label{fig:varyingLp}
\end{figure}

Figure \ref{fig:cp2pd} summarises our basic theoretical expectations for the (1+1)-d $\mathbb{C}$P(2) model $\mu_3$-$\mu_8$ phase diagram at zero temperature. A definite expectation is that the system has a mass gap $m$. Hence our first goal was to confirm this: Figure \ref{fig:varyingLp} shows results for the particle number density $\langle n\rangle$, for simulations with $\beta J=140.75$, $L/a=250$, $\epsilon J=0.05$, $L'/a=8,10,12$, $\mu_8=0$ and $\mu_3>0$. From measurements of the correlation length at zero chemical potential, the rest energy gaps in each $L'$ case were calculated to be $mc^2=0.1248(10)J$, $0.0623(2)J$, and $0.0322(3)J$, respectively. 
The error is relatively small for $L'/a=10$ because more runs were performed in this case. As we can see in Figure \ref{fig:varyingLp}, $\langle n\rangle$ starts to become significantly larger than zero close to the point where $\mu_3=mc^2$, agreeing with expectation. 

\begin{figure}
\centering
\begin{tabular}{cc}
\hspace{-35pt}\includegraphics[scale=0.8, trim = 0 0 65 0, clip=true]{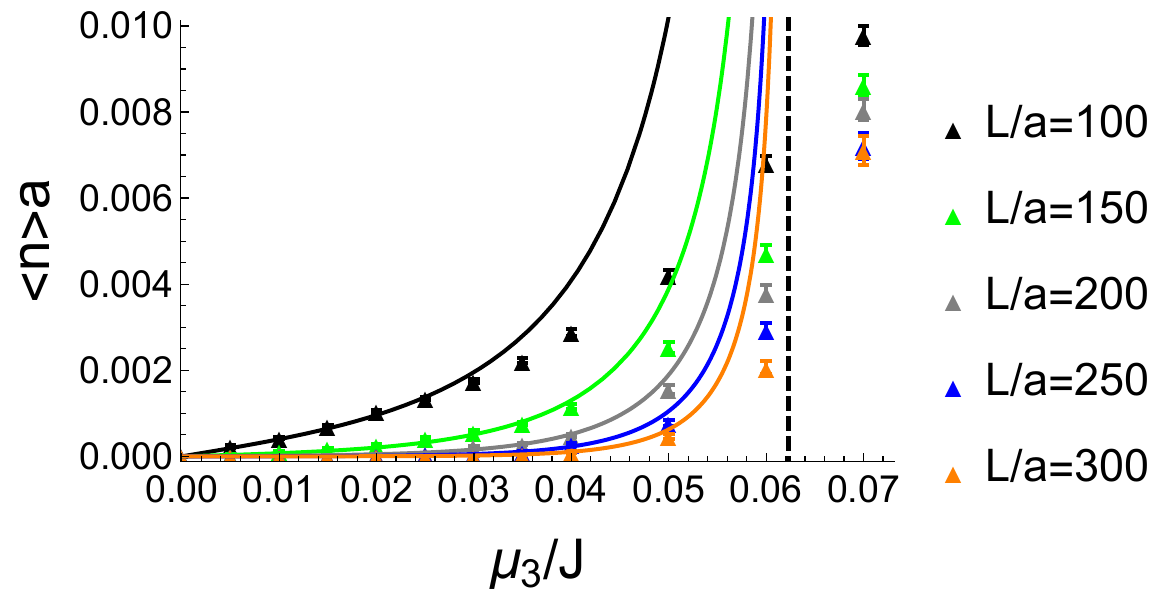} & \includegraphics[scale=0.8, trim = 20 0 0 0, clip=true]{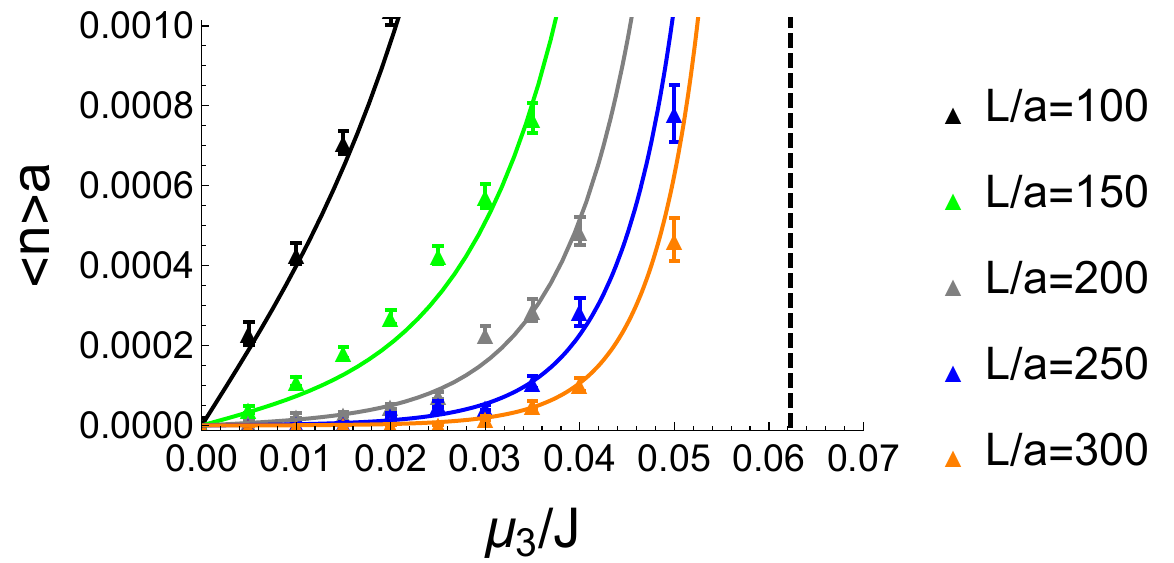} \\
\hspace{-8pt}(a) & \hspace{-38pt}(b)
\end{tabular}
\caption{Particle number density plotted against chemical potential for $L'=10$, $\epsilon J=0.05$ and various $L$, where $L\approx\beta c$. (b) is simply a zoom-in of (a). The solid lines show the prediction of the free boson model for $\langle n \rangle$. The free boson model is formulated in the continuum rather on the lattice, hence there are an infinite number of degrees of freedom available, this permits an infinite number of bosons to be produced once $\mu_3=mc^2$ is reached, at this point $\langle n \rangle\rightarrow\infty$ since there are no interactions to limit production.}
\label{fig:thermLim}
\end{figure}

For our finite system a sharp increase in $\langle n\rangle$ precisely at $\mu_3=mc^2$ is not observed, we suspect this is a finite size effect. In an infinite system with an associated infinite number of degrees of freedom, we would expect there to be a non-analytic point at $\mu_3=mc^2$, with $\langle n\rangle=0$ for $\mu_3<mc^2$, and $\langle n\rangle > 0$ for $\mu_3>mc^2$. In agreement with this expectation, Figure \ref{fig:thermLim} shows that for $\mu_3<mc^2$, $\langle n\rangle$ tends towards zero, for any given value of $\mu_3$, as the thermodynamic limit is approached. Here we also compare the results with a simple model for the particle number density of a free boson gas,
\begin{equation}
\langle n\rangle=\frac{\langle N^+\rangle-\langle N^-\rangle}{L}, \ \ \ \ \langle N^\pm\rangle=\sum_{p}\frac{1}{{\rm e}^{\beta\left(\sqrt{p^2c^2+m^2c^4}\mp\mu\right)}-1},
\end{equation}
where $p=2\pi l/L$, $l\in \mathbb{Z}$, are the allowed momenta in the (1+1)-d system. Note that the allowed momenta do not form a finite set because the model assumes a spatial continuum rather than lattice, we did not undertake the significant effort to calculate the lattice dispersion relation for the model because at low energies we expect the continuum dispersion relation to be very similar. For $0<\mu_3\ll mc^2$  we see good agreement between the free boson model and numerical data, this makes sense because at very small values of $\mu_3$ we have a system so dilute that it most often consists of zero or one particle, in this case particle interactions are negligible. For larger values of $\mu_3$ interactions between bosons become significant and we no longer expect agreement between the numerical data and the free boson model.

\begin{figure}
\centering
\begin{tabular}{cc}
\includegraphics[scale=0.55, trim = 0 0 0 0, clip=true]{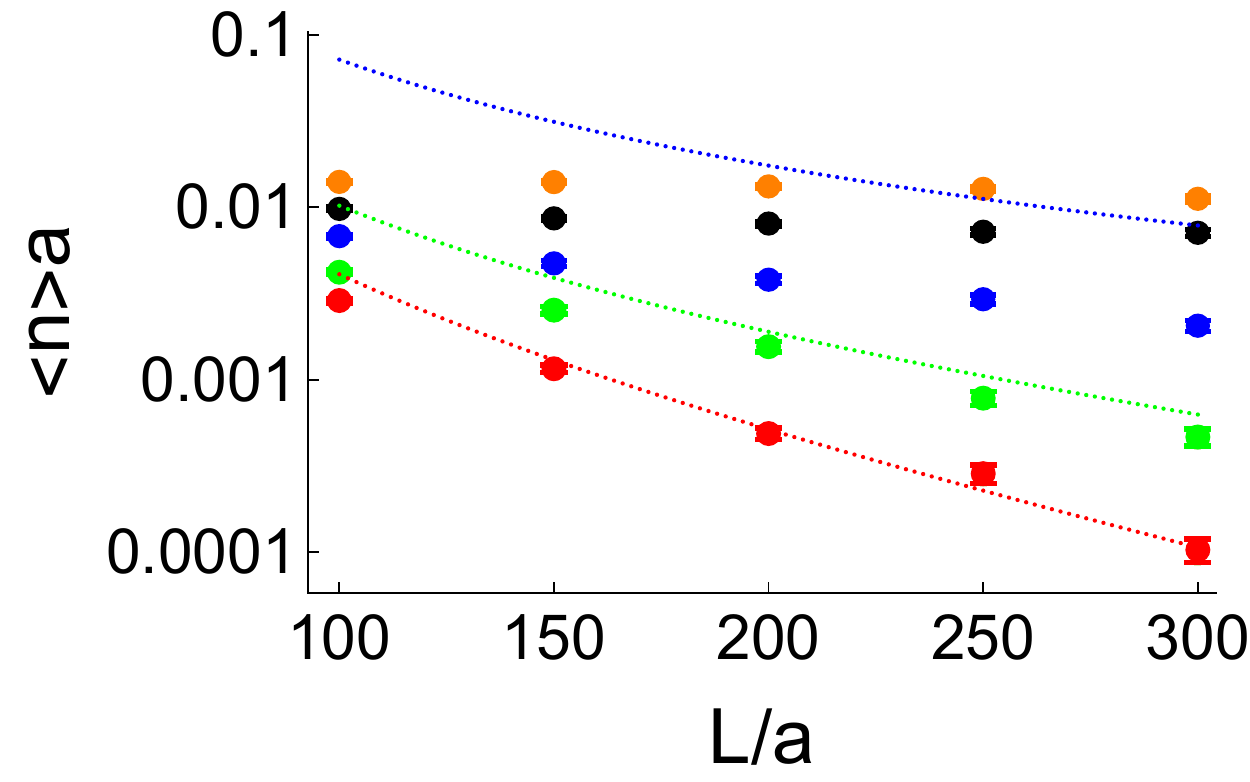} & \includegraphics[scale=0.55, trim = 0 0 0 0, clip=true]{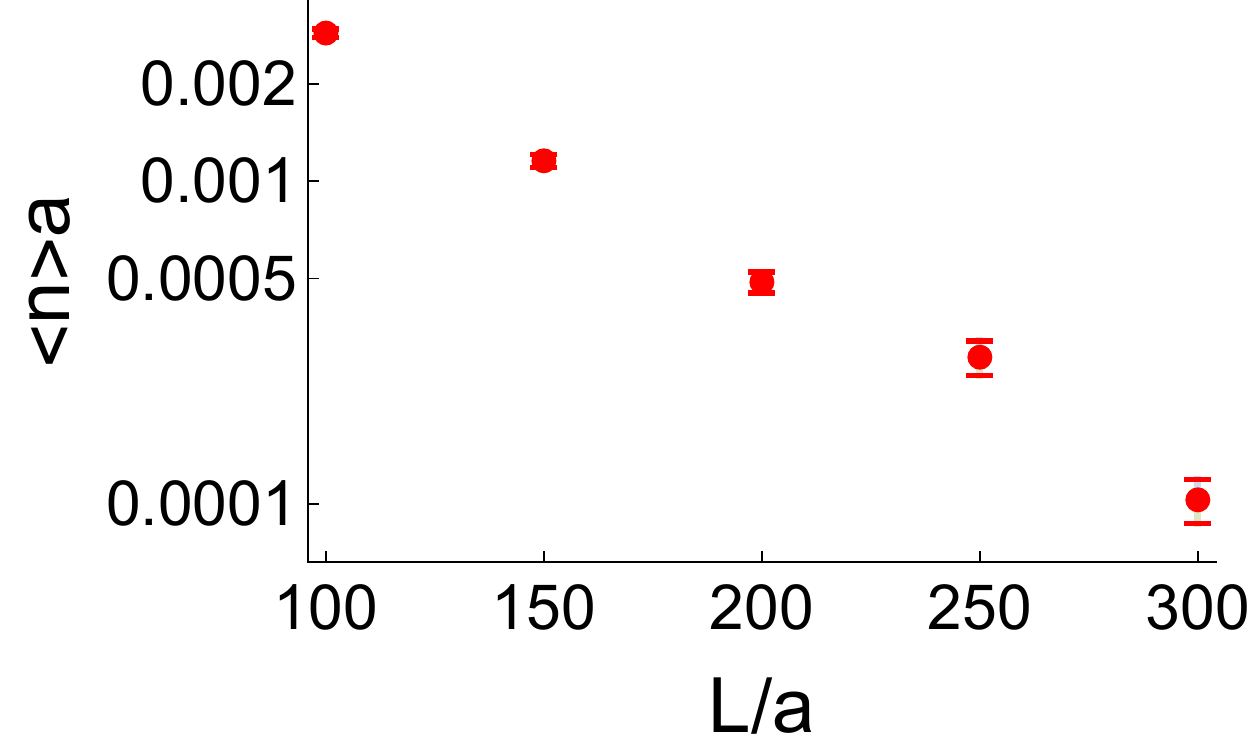}  \\
\hspace{1.35cm} (a) & \hspace{1.55cm} (b) \\
& \\
\includegraphics[scale=0.55, trim = 0 0 0 0, clip=true]{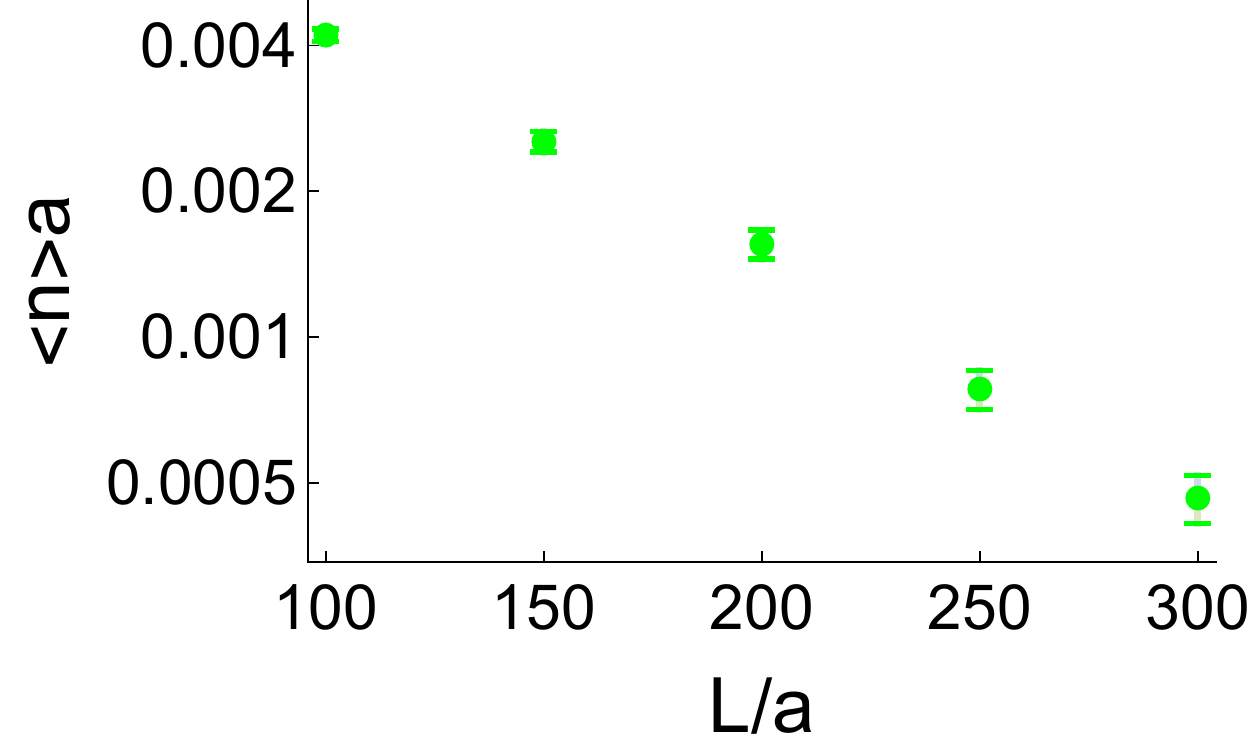} & \includegraphics[scale=0.55, trim = 0 0 0 0, clip=true]{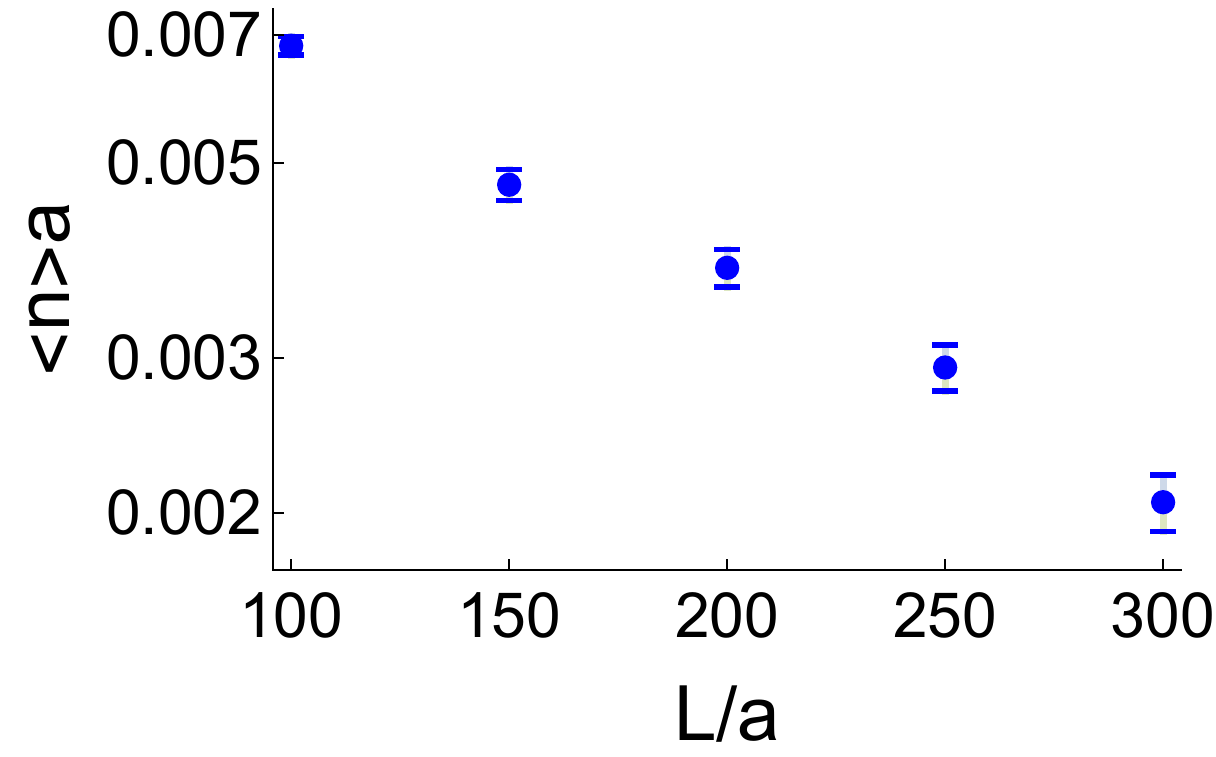} \\
\hspace{1.5cm} (c) & \hspace{1.2cm} (d) \\
& \\
\includegraphics[scale=0.55, trim = 0 0 0 0, clip=true]{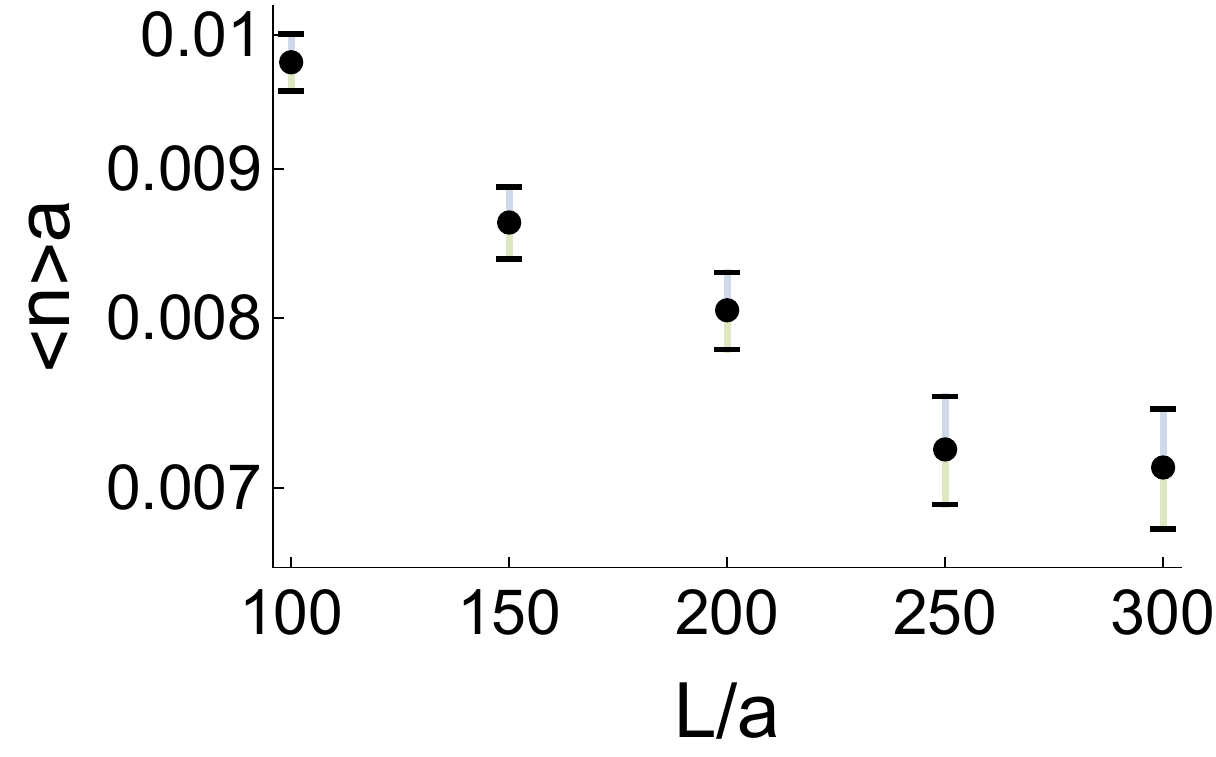} & \includegraphics[scale=0.55, trim = 0 0 0 0, clip=true]{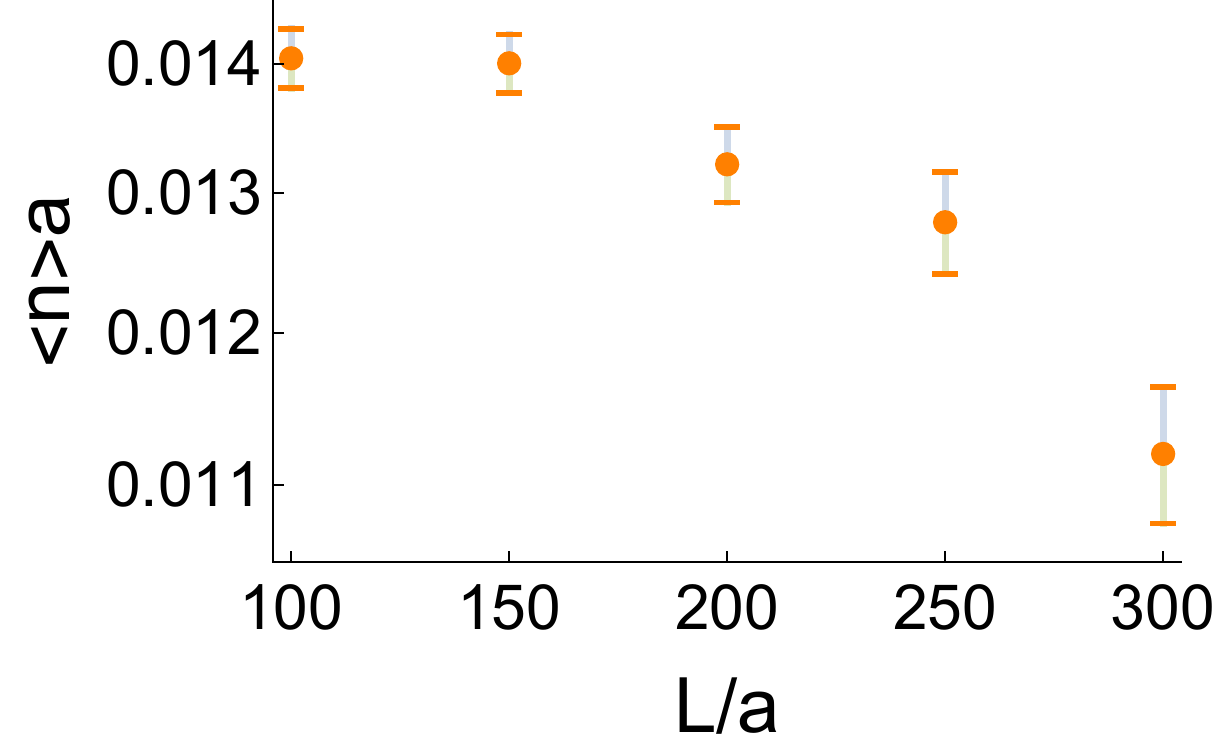}  \\
\hspace{1.2cm} (e) & \hspace{1.2cm} (f) \\
\end{tabular}
\caption{Particle number density plotted on a logarithmic scale against system size for $L\approx\beta c$, $L'/a=10$, $\epsilon J=0.05$ and the $\mu_3$ values: (b) 0.04$J$ (red), (c) 0.05$J$ (green), (d) 0.06$J$ (blue), (e) 0.07$J$ (black), (f) 0.08$J$ (orange). In (a), (b)-(f) are combined, the dotted lines are the expectations from the free boson gas model for $\mu_3$: 0.04$J$ (red), 0.05$J$ (green), 0.06$J$ (blue), although we expect exponential scaling for these values of $\mu_3$ since they are below $mc^2=0.0623(2)J$, and hence straight line behaviour on a log-plot, we find it interesting to show from this perspective how the boson gas model increasingly agrees with the data as the thermodynamic limit is taken, but decreasingly as $\mu_3$ increases towards $mc^2$.}
\label{fig:scaling}
\end{figure}

Although the free boson gas model can not help us check the numerical results above and near $\mu_3=mc^2$, we can check the data in this region by studying how $\langle n\rangle$ scales for particular values of $\mu_3$ in the thermodynamic limit. The particle density is suppressed by the Boltzmann factor $e^{-\beta(mc^2-\mu_3)}$, hence, for increasing $\beta$, we can expect $\langle n \rangle\rightarrow0$ exponentially slowly for $\mu_3<mc^2$, but for $\mu_3>mc^2$ the scaling is not clear and unlikely to be exponential since there will be many particles interacting strongly. Figures \ref{fig:scaling} (b)-(d) confirm that for $\mu_3<mc^2$ the particle number density does indeed scale exponentially. Figures \ref{fig:scaling} (e)-(f) do not show such clear exponential behaviour, as we expected, but it is difficult to say that another particular type of scaling is being followed since (e) and (f) do not look similar; perhaps in (d)-(f) we are seeing the transition from exponential to another type of scaling or perhaps because the uncertainties are more significant in these cases the central values have landed in deceiving positions. At the very least we have not found evidence which contradicts the validity of the theoretical framework.

\vspace{-8pt}
\section{Conclusions and Outlook}
\vspace{-8pt}
Using D-theory, the (1+1)-d $\mathbb{C}$P(2) model has been simulated at non-zero chemical potential, and the particle number density is observed to behave in a manner congruent with expectations. Specifically: i) $\langle n\rangle$ becomes significantly larger than zero beyond the value of the chemical potential equal to the system's expected rest energy gap; ii) at low densities the results agree with a free boson gas model remarkably well, above and near $\mu_3=mc^2$ the density is larger and we do not expect there to be agreement since the simple free boson gas model is inadequate; iii) for $\mu_3<mc^2$ the results for $\langle n\rangle$ scaling with temperature exhibit the expected exponential scaling, while for $\mu_3>mc^2$ we have some evidence that another type of scaling is present, which would agree with expectations.

In work to be published soon the study presented here will be repeated for open boundary conditions in the $L'$-direction to allow direct comparison with future experiments. Furthermore simulations for $\mu_8\neq 0$ and measurements of spin-spin correlation functions will be included to permit a more comprehensive expostion of the $\mathbb{C}$P(2) $\mu_3$-$\mu_8$ phase diagram.
\vspace{-8pt}
\section*{Acknowledgements}
\vspace{-8pt}
This work was supported by the Schweizerischer Nationalfonds, the European Research Council by means of the European Union's Seventh Framework Programme (FP7/2007- 2013)/ERC grant agreement 339220, and the Mexican Consejo Nacional de Ciencia y Tecnolog\'ia (CONACYT) through projects CB-2010/155905 and CB-2013/222812, and by DGAPA-UNAM, grant IN107915.


\vspace{-8pt}
\begin{thebibliography}{99}
\vspace{-8pt}
\bibitem{Nascimbene12}
I.~Bloch, J.~Dalibard, S.~Nascimb{\`e}ne, Nat. Phys. 8~(4) (2012) 267--276.
\vspace{-5pt}
\bibitem{Laflamme:2015wma}
  C.~Laflamme, W.~Evans, M.~Dalmonte, U.~Gerber, H.~Mej\'ia-D\'iaz, W.~Bietenholz, U.-J.~Wiese, P.~Zoller, Annals Phys. 370 (2016) 117.
\vspace{-5pt}
\bibitem{Laflamme:2015pos}
  C.~Laflamme, W.~Evans, M.~Dalmonte, U.~Gerber, H.~Mej\'ia-D\'iaz, W.~Bietenholz, U.-J.~Wiese, P.~Zoller, PoS(LATTICE2015) 311.
\vspace{-5pt}
\bibitem{Brower:1999kq}
R.~Brower, S.~Chandrasekharan, U.-J. Wiese, Phys. Rev. D {\bf 60}~(9) (1999) 094502.
\vspace{-5pt}
\bibitem{Brower2004149}
R.~Brower, S.~Chandrasekharan, S.~Riederer, U.-J. Wiese, Nucl. Phys. B {\bf 693}~(1--3) (2004) 149 -- 175.
\vspace{-5pt}
\bibitem{Beard:2004jr}
  B.~B.~Beard, M.~Pepe, S.~Riederer, U.~J.~Wiese,
  Phys.\ Rev.\ Lett.\  {\bf 94} (2005) 010603.
\vspace{-5pt}
\bibitem{Sachdev89}
N.~Read, S.~Sachdev, Phys. Rev. Lett. {\bf 62} (1989) 1694--1697.
\vspace{-5pt}
\bibitem{Harada:2003}
  K.~Harada, N.~Kawashima, M.~Troyer,
  Phys.\ Rev.\ Lett.\ {\bf 90} 117203.
\vspace{-5pt}
\bibitem{Beard:2006mq}
  B.~B.~Beard, M.~Pepe, S.~Riederer, U.-J.~Wiese,
  Comput.\ Phys.\ Commun.\  {\bf 175} (2006) 629.
\vspace{-5pt}
\bibitem{Evertz2003}
H.~G.~Evertz, Adv. Phys. 52:1, 2003.
\vspace{-5pt}
\bibitem{Sandvik2002}
O.~F.~Syljuasen, A.~W.~Sandvik, Phys. Rev. E {\bf 66}, (2002) 046701.
\vspace{-5pt}
\bibitem{Jiang:2010cd}
  F.-J.~Jiang and U.-J.~Wiese, Phys. Rev. B {\bf 83}, (2011) 155120.


  

\end{thebibliography}
\end{document}